\documentclass[conference]{IEEEtran}
\IEEEoverridecommandlockouts
\usepackage{cite} 
\usepackage{array}
\usepackage{multirow}
\usepackage{amsmath,amssymb,amsfonts}
\usepackage{graphicx}
\usepackage{textcomp}
\usepackage{algorithm, algpseudocode}
\usepackage{xcolor}
\usepackage{soul}
\usepackage{algorithm}
\usepackage{algpseudocode}
\def\BibTeX{{\rm B\kern-.05em{\sc i\kern-.025em b}\kern-.08em
    T\kern-.1667em\lower.7ex\hbox{E}\kern-.125emX}}
\begin{document}

\title{AI-driven Reverse Engineering of QML Models}

\author{\IEEEauthorblockN{Archisman Ghosh}
\IEEEauthorblockA{\textit{CSE Department} \\
\textit{Penn State University}\\
State College, PA, USA\\
apg6127@psu.edu}
\and
\IEEEauthorblockN{Swaroop Ghosh}
\IEEEauthorblockA{\textit{School of EECS} \\
\textit{Penn State University}\\
State College, PA, USA \\
szg212@psu.edu}}


\maketitle

\begin{abstract}

Quantum machine learning (QML) is a rapidly emerging area of research, driven by the capabilities of Noisy Intermediate-Scale Quantum (NISQ) devices. With the progress in the research of QML models, there is a rise in third-party quantum cloud services to cater to the increasing demand for resources. New security concerns surface, specifically regarding the protection of intellectual property (IP) from untrustworthy service providers. One of the most pressing risks is the potential for reverse engineering (RE) by malicious actors who may steal proprietary quantum IPs such as trained parameters and QML architecture, modify them to remove additional watermarks or signatures and re-transpile them for other quantum hardware. Prior work presents a brute force approach to RE the QML parameters which takes exponential time overhead. In this paper, we introduce an autoencoder-based approach to extract the parameters from transpiled QML models deployed on untrusted third-party vendors. We experiment on multi-qubit classifiers and note that they can be reverse-engineered under restricted conditions with a mean error of order $10^{-1}$. The amount of time taken to prepare the dataset and train the model to reverse engineer the QML circuit being of the order $10^3$ seconds (which is $10^2 \times$ better than the previously reported value for 4-layered 4-qubit classifiers) makes the threat of RE highly potent, underscoring the need for continued development of effective defenses.

\end{abstract}


\section{Introduction} \label{intro}
Quantum Machine Learning (QML) combines quantum computing and artificial intelligence to harness the unique capabilities of quantum systems and solve complex problems far beyond the reach of classical machine learning techniques. It can offer exponential speedups in tasks such as optimization, data classification, and pattern recognition \cite{schuld2015introduction}. As the quantum computing landscape advances, particularly with the advent of Noisy Intermediate-Scale Quantum (NISQ) devices \cite{Preskill_2018}, the potential for QML applications in various industries is rapidly expanding. This growth is leading to a surge in third-party quantum cloud service providers, offering access to various price and performance points. However, this shift also introduces several significant security challenges, particularly regarding the protection of intellectual property (IP) and sensitive data in QML circuits. As more organizations rely on external quantum hardware providers, the risk of IP theft, committed by untrusted third-party cloud providers or rogue adversaries sharing the same resources with unsuspecting users, including the unauthorized replication and reverse engineering (RE) of QML models, becomes a pressing concern \cite{suryansh_acm}.

    
\subsection{Why Protect QML Models}
QML models are particularly vulnerable to security risks due to several factors inherent to their development and deployment. First, QML training is extremely costly. The cost of hosting on quantum processing units (IBM's Heron r2 QPU with 156 qubits cost a user \$1.60/s) is almost $\sim 10^4 \times$ more than classical GPU platforms (Google Cloud Platform rates NVIDIA V100 GPUs at \$2.51/hr). The high cost and time-intensive nature of training QML models make them valuable targets for adversaries seeking to bypass these expenses through illicit means. The reliance on third-party quantum cloud providers, which often lack stringent security protocols, further exacerbates the risk. Since these providers handle the quantum hardware and data, any breach in their security could lead to unauthorized access to sensitive QML models. Additionally, QML models often encapsulate numerous intellectual properties (IPs), from unique quantum circuit designs to proprietary training data and algorithms, making them rich targets for IP theft. The combination of high development costs, reliance on potentially untrusted cloud services, and the concentration of valuable IPs within QML models creates a perfect target for security threats, making it imperative to develop robust protection mechanisms to safeguard these cutting-edge technologies.

\subsection{Attack Model and Motivation}
\begin{figure}
    
    \centering
    \includegraphics[width=1\linewidth]{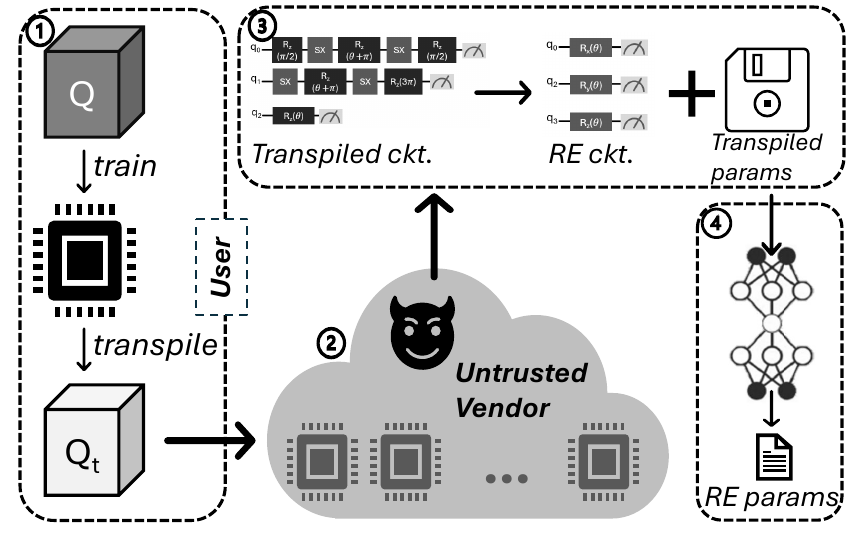}
    \caption{The threat model of reverse engineering user-designed QML models by untrusted vendors. (1) shows the training and transpilation of a QML model $Q$ in non-proprietary quantum hardware; (2) shows the deployment of the trained QML model $Q_t$ on a cloud service provided by an untrusted vendor; (3) demonstrates the reverse engineering of the transpiled QML circuit to the RE circuit and the transpiled params. This can be done by an adversary with the help of pre-designed LUTs \cite{ghosh2024quantumimitationgamereverse}; (4) shows the procedure of feeding the transpiled parameters into an autoencoder to generate the reverse-engineered parameters by the untrusted vendor.}
    \label{fig:flow}
    \vspace{-10pt}
\end{figure}

The user embeds the input data into the trained QML model, transpiles it to optimize the circuit for a target quantum hardware, and sends it to the quantum cloud provider for execution. Access to the white-box architecture of this trained QML circuit could enable untrusted cloud providers to steal and misuse it (Fig. \ref{fig:flow}). For instance, an adversary could remove the state preparation circuit, extract the trained portion of the quantum neural network (QNN) in the transpiled form (Fig. \ref{fig:flow} (3)), and reverse engineer the QNN architecture as well as the trained parameters (Fig. \ref{fig:flow} (4)), and use their own input data for inference on the same hardware. They could also sell the trained QNN. Prior knowledge of the original circuit design and parameters is not necessary for such attacks. However, access to these details using RE could offer additional advantages, such as adapting the model for different hardware platforms, revealing the entanglement architecture of the QNN for cloning or resale, and tampering with the model, including watermark alteration or embedding.

\subsubsection{RE of QML model}
Reverse engineering involves analyzing a model to recreate its design while preserving its architectural details \cite{re1}. In the context of quantum computing, this process entails reconstructing the original hardware-agnostic quantum circuit from its optimized, hardware-specific transpiled form. 

\subsubsection{Challenges in RE}
Reverse engineering QML models poses distinct challenges compared to classical machine learning models due to differences in representation, transpilation, and hardware dependencies. Classical ML models are typically represented as mathematical functions or neural networks, whereas QML models are expressed as quantum circuits with quantum gates serving as parameters. The transpilation process is crucial in adapting the QML circuit to the native gate set of the specific quantum hardware used for training and inference.

Although identifying the entanglement architecture from a transpiled quantum circuit is relatively straightforward—requiring the reversal of logical-to-physical qubit mapping and accounting for $SWAP$ gates—recovering the original parameters from decomposed and optimized single-qubit gates is significantly more complex. The challenges arise from several factors: (i) Transpilation converts all single-qubit gates into basis gates (e.g., $RZ(\theta)$) on IBM machines), making it difficult to identify the original type of rotation gates. (ii) Parameterized rotation gates are transpiled into sequences of single-qubit gates that are further optimized with rotation gates derived from other parts of the QML model, complicating parameter recovery. (iii) The transpilation process often introduces a global phase to maintain correct relative phases between qubit states, further obscuring the original parameters. (iv) Increasing the optimization level during transpilation enforces stricter optimization rules, adding a layer of complexity to the reverse engineering process.

\subsection{Contributions}
Although the problem of RE of a QML model has been presented \cite{ghosh2024quantumimitationgamereverse}, the approach involves brute force search for parameters. \emph{To the best of our knowledge, this is the first effort to employ machine learning techniques to perform RE of a QML circuit.} The major contributions are as follows:

\begin{enumerate}

    \item We propose an autoencoder-based procedure to extract the original parameters from transpiled QML circuits.
    \item We demonstrate the efficacy of the proposed idea by reverse engineering multi-qubit classifiers.
    \item We perform a detailed comparison with existing techniques and discuss security concerns.
\end{enumerate}

\subsection{Paper Structure}
Section II provides a background and related works. Section III covers the threat model. Section IV presents the proposed machine learning-based RE approach and Section V performs a comparative analysis with the state-of-the-art. Section VI concludes the paper.
\section{Background}
\label{bg}

\subsection{Quantum Computing}
In quantum computing, the qubit, or quantum bit, is the fundamental unit of information. Unlike classical bits, which can only be in one of two states—0 or 1—a qubit can be in a superposition of both states simultaneously. In Dirac notation, the state of a qubit is denoted as $|\psi\rangle = \alpha|0\rangle + \beta|1\rangle$, where $\alpha$ and $\beta$ are complex numbers that must satisfy the normalization condition $|\alpha|^2 + |\beta|^2 = 1$. The computational basis states are represented as $|0\rangle = [1 \ 0]^T$ and $|1\rangle = [0 \ 1]^T$. When dealing with multiple qubits, $n$ qubits can represent a quantum state within a $2^n$-dimensional space, with basis states ranging from $|0\dots0\rangle$ to $|1\dots1\rangle$. An $n$-qubit quantum state can be expressed as $|\psi_n\rangle = \sum_{i=0}^{2^n-1} a_i |i\rangle$, where the coefficients $a_i$ satisfy the normalization condition $\sum_{i=0}^{2^n-1} |a_i|^2 = 1$.

\subsection{Compilation of Quantum Circuits}
Compilation of quantum circuits involves translating high-level quantum programs into a form compatible with the constraints of quantum hardware, a process known as \textit{transpilation} in IBM terminology \cite{trans}. The following steps are required in the transpilation process. \textbf{Gate Translation:} Quantum programs are typically written using high-level gates, which represent abstract quantum operations. However, current quantum computers support only a limited set of native instructions, or basis gates, such as \texttt{[id, x, sx, cnot, rz]} on IBM machines. High-level instructions must be translated into these basis gates to execute on the hardware, aligning the quantum algorithm with hardware limitations. \textbf{Coupling Map Constraints:} The coupling map constraint arises from the physical layout of qubits in the quantum hardware. For instance, for two-qubit gates like $CNOT$, the two qubits involved must be physically connected to each other. In case the qubits are not physically connected, a $SWAP$ operation is performed to interchange states between the physical qubits and perform the $CNOT$ operation. \textbf{Optimization: }This step involves combining multiple single-qubit gates, canceling, and even reordering to reduce the complexity of the circuit. This process also helps in faster execution. 

\subsection{Quantum Neural Networks}
Quantum Neural Networks (QNNs) combine quantum computing with machine learning \cite{Schuld2014}. By designing quantum circuits to embed classical data as qubit states, QNNs can perform tasks such as regression and classification, similar to classical neural networks. The following steps are required in the QNN design. \textbf{Quantum Data Encoding:} This step involves embedding classical data into the Hilbert space using quantum states. Techniques include amplitude encoding, which normalizes data into qubit amplitudes; angle encoding, which converts data into rotation angles for qubits; and basis encoding, which maps binary data to computational basis states. \textbf{Parameterized Quantum Circuits (PQC):} PQCs comprising tunable quantum gates form the core component of any QNN. PQCs feature single-qubit rotation gates like $RX(\theta), RY(\theta)$, and $RZ(\theta)$, and multi-qubit rotation gates like $CRX(\phi), CRY(\phi)$, and $CRZ(\phi)$ with adjustable parameters that define qubit interactions. Entanglement between qubits enhances computational capabilities, allowing PQCs to perform complex transformations, similar to layers in classical neural networks. \textbf{Measurement:} After computation, measurement extracts classical information from quantum states, collapsing them to reveal the final qubit states (0 or 1). The probabilities of each basis state are measured to derive outputs, which are then processed classically. The QNN workflow starts with preprocessing and encoding classical data into quantum states, followed by processing within the PQC. Classical optimization algorithms iteratively adjust PQC parameters by minimizing the loss function until the QNN converges.

\subsection{Related Work}

RE attacks on convolutional neural networks, running on hardware accelerators, using side-channel attacks to infer the network structure and even extract CNN weights despite data encryption have been studied \cite{revcnn}. Fault attacks have also been used to reverse engineer neural networks \cite{Breier_2022}. 
In classical domains, black-box neural networks can be attacked by querying the model and observing its outputs, and training a metamodel on them to predict the original architecture \cite{Oh2019}. 

The extension of RE to quantum circuits is relatively unexplored. There have been attempts to RE the parameters of QML circuits \cite{ghosh2024quantumimitationgamereverse} from the transpiled circuit. The adversary compiles a LUT based on the possible ordering of rotation gates and the corresponding transpiled forms, parses the transpiled circuit, and compares the ordering of rotation gates and entanglement from the LUT to reverse-engineer the architecture of the QML circuit. The parameters of the rotation gates can also be brute forced out. Fig. \ref{fig:re} shows a subsection of the LUT from \cite{ghosh2024quantumimitationgamereverse}. The adversary parses the transpiled QNN to obtain the ordering of rotation gates (left in Fig. \ref{fig:re}). Using this ordering, the corresponding reverse-engineered order of gates is obtained from the LUT. For Fig. \ref{fig:re}(1) case, an order of $RZ(\theta_0) \cdot SX \cdot RZ(\theta_1) \cdot SX \cdot RZ(\theta_2)$ can be reverse engineered as an order of $RX(\phi_0) \cdot RY(\phi_1) \cdot RZ(\phi_2)$. Further, the adversary calculates the values of reverse-engineered parameters ($\phi$) from the original parameters ($\theta$) using brute force. While obtaining the rotation gates from the transpiled circuit is novel, extracting parameters consumes significant time ($\sim 10^6$ for 8-layered, 4-qubit classifiers) undermining the threat to QML models hosted in untrusted environments. Blind computation \cite{blind} can significantly increase communication overhead due to the need for additional quantum states, operations, and qubits. Moreover, it is vulnerable to decoherence errors and background noise in quantum hardware.

\section{Threat Model and Analysis}

\subsection{Threat Model}
We assume that a malicious insider in the quantum cloud provider may try to gain deeper insights into the victim model motivated by the fact that QMLs trained on non-proprietary hardware are costly in terms of the training weights and circuit design. The adversary might even try to make a profit by offering services with the stolen model. With access to a transpiled QML circuit, the adversary can strip the state preparation circuit, parse the transpiled QNN architecture, and compare it with standard transpilation forms to obtain a close copy of the user model. The trained weights (parameters) can also be reverse-engineered by repeated transpilation of the copied circuit on a simulator. The adversary could attach the reverse-engineered circuit to a custom state preparation circuit and execute it on the same hardware for a different dataset. With access to the reverse-engineered QML circuit, the adversary could (i) transpile the model for different quantum hardware and qubit technologies, increasing the marketability of the stolen model, (ii) avoid legal issues by tampering with or removing any embedded watermarks or embedding their own, and (iii) further train the model for specific applications. In this paper, we propose autoencoders to learn the complex mathematical model of the transpilation of rotation gates under specific optimization conditions. The goal of the adversary is to predict the original parameters so that they match that of the reverse-engineered circuit closely.

\subsection{Adversary Capabilities}
We assume that the untrusted third-party quantum hardware provider has the following capabilities: (i) white-box access to the transpiled QML circuit, which serves as an input to the reverse-engineering autoencoder model and helps in predicting the original parameterized rotation gates; (ii) access to the transpiler, allowing them to transpile the reverse-engineered version of the model and check the accuracy of their predictions; and (iii) substantial computational resources to prepare the dataset, train the autoencoder model, and expedite the search for parameters, minimizing the error between the original and reverse-engineered QML models.
\section{Proposed Idea}

\begin{figure}
    
    \centering
    \includegraphics[width=\linewidth]{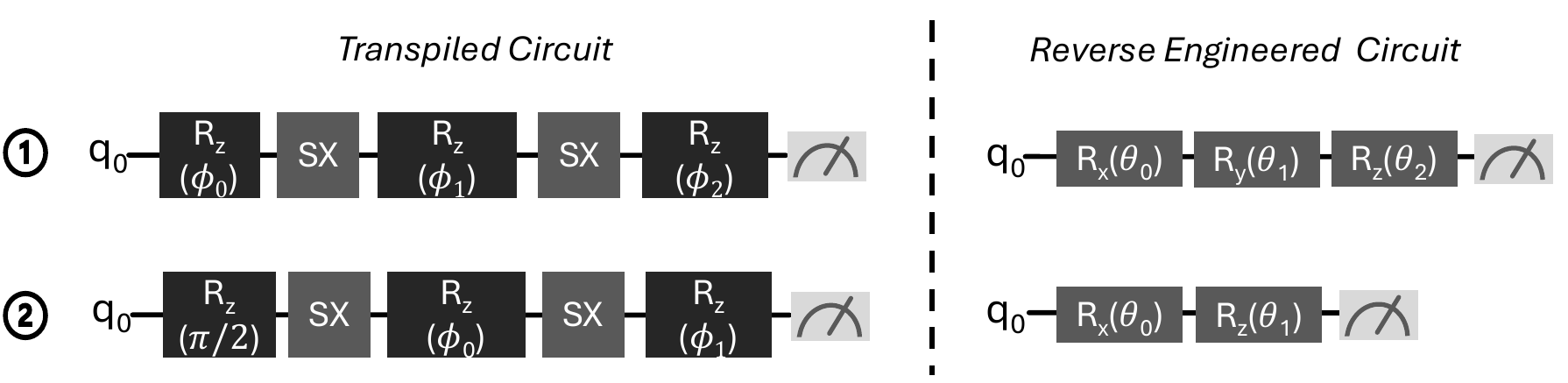}
    \caption{A subset of the LUT discussed in \cite{ghosh2024quantumimitationgamereverse}. The adversary obtains the transpiled circuit, parses it qubit-by-qubit, and obtains an ordering of rotation gates which is matched with the LUT and the original order of gates is obtained. Following this, the adversary feeds the transpiled parameters into the autoencoder and predicts the parameters as close as possible to the original parameters thus effectively reverse engineering the QML model.}
    \label{fig:re}
    \vspace{-10pt}
\end{figure}
\subsection{Reverse Engineering}
We advance the procedure to reverse engineer the parameters of a QML circuit substantially compared to  \cite{ghosh2024quantumimitationgamereverse} in this work. A QML circuit gets converted into a transpiled form using the native gate set of the quantum hardware. The authors in \cite{ghosh2024quantumimitationgamereverse} prepare a LUT based on the typical patterns observed during the transpilation procedure in specific hardware under certain conditions. The adversary can use such an LUT to determine sections of the QML circuit like the order of rotation gates and possible entanglement architecture. The transpiled circuit can be parsed qubit-by-qubit and the order of gates and position of $CNOT$ gates can help to determine the original circuit structure. On obtaining the order of rotation gates, the adversary can brute-force the original parameters by comparing the transpiled versions of the original circuit and the reverse-engineered circuit over $[-\pi, \pi]$. The brute-force approach, although relatively accurate, is extremely slow, taking time in the order of $10^6$ seconds for 3-layered 8-qubit classifiers. In contrast, we propose an autoencoder-based approach for the extraction of parameters. Since the transpilation process in quantum circuits is a mathematical transformation, the parameters in the original and transpiled circuit can be described as a one-to-one mapping and hence be learned by an autoencoder to predict the potential rotations of the original circuit given the transpiled parameters. In the following subsections, we discuss the design of the individual components of the proposed autoencoder architecture and the dataset used for training.  

\subsection{Dataset}
We prepare the dataset based on the requirements of the autoencoder. For example, in Fig. \ref{fig:re}(1), we can observe that the template transpiled circuit for the adversary has three parameters. Therefore we train the autoencoder model on a dataset comprising all possible combinations in the form $(x, y, z)$ from the set of values in $[-\pi, \pi]$, separated by a step size of 0.1 as inputs and the corresponding set of  $RZ(\theta)$ parameters in the transpiled circuit as the output. In Fig. \ref{fig:re}(2), we see that the template circuit for the adversary has two parameters and hence the dataset is prepared in the same fashion as above but the combinations in $[-\pi,\pi]$ is done in the form of $(x, y)$. We train the input set on the encoder and obtain the predicted parameters from the decoder.

\begin{table}[h!]
\centering
\caption{Encoder Architecture}
\begin{tabular}{l|c|c|c}

\textbf{Layer Type}       & \textbf{Output Shape} & \textbf{Activation} & \textbf{\# Params} \\ \hline \hline

Input Layer               & \# transpiled params                     & -                            & 0                      \\ 
Dense Layer 1             & 256                   & ReLU                         & 1024                   \\ 
Batch Norm 1              & 256                   & -                            & 1024                   \\ 
Dropout 1                 & 256                   & -                            & 0                      \\
Dense Layer 2             & 128                   & ReLU                         & 32896                  \\
Batch Norm 2              & 128                   & -                            & 512                    \\ 
Dropout 2                 & 128                   & -                            & 0                      \\ 
Dense Layer 3             & 64                    & ReLU                         & 8256                   \\ 
Dense Layer 4             & 32                    & ReLU                         & 2080                   \\ 
Latent Space              & 16                    & ReLU                         & 528                    \\ \hline \hline
\end{tabular}

\label{tab:encoder_model}
\vspace{-15pt}
\end{table}
\subsection{Encoder and Decoder}
The encoder is designed (Table \ref{tab:encoder_model}) to compress $k$-dimensional input data into a lower-dimensional latent space, $k$ being the number of parameters in the rotation gates of the transpiled circuit. The value of $k$ in Fig. \ref{fig:re}(1) is 3. The model begins with an input layer, followed by dense layers with 256, 128, 64, and 32 neurons, each using ReLU activation for non-linearity. Batch normalization is applied after certain layers to stabilize learning, and dropout layers (30\%) are included to prevent overfitting since the transpilation process is a highly non-linear mathematical operation. The final layer of the encoder reduces the input data to a 16-dimensional latent representation, effectively capturing the most essential features of the original. The decoder is designed (Table \ref{tab:decoder_model}) to reconstruct the original input from the 16-dimensional latent space. It starts with an input layer that accepts the latent vector and passes it through dense layers with 32, 64, 128, and 256 neurons, each with ReLU activation. Batch normalization and dropout layers are used similarly to the encoder to ensure stability and prevent overfitting. The final layer of the decoder outputs a $k$-dimensional vector, reconstructing the original input data from the compressed latent representation provided by the encoder.

\subsection{Training}
The autoencoder combines the encoder and decoder, forming a complete model that maps $k$-dimensional input data to a compressed latent space and then reconstructs it. The model is compiled with the Adam optimizer and uses Mean Squared Error (MSE) as the loss function, aiming to minimize the reconstruction error. Training involves feeding the autoencoder, outputs from the dataset made, i.e., the transpiled parameters, and learning to map these parameters back to the original input combinations (the combination of parameters over $[-\pi,\pi]$). The model is trained over 100 epochs with a batch size of 1024, using 20\% of the data for validation.

\begin{table}[h!]
\centering
\caption{Decoder Architecture}
\begin{tabular}{l|c|c|c}

\textbf{Layer Type}       & \textbf{Output Shape} & \textbf{Activation} & \textbf{\# Params} \\ \hline \hline

Latent Input              & 16                    & -                            & 0                      \\ 
Dense Layer 1             & 32                    & ReLU                         & 544                    \\ 
Batch Norm 1              & 32                    & -                            & 128                    \\ 
Dropout 1                 & 32                    & -                            & 0                      \\ 
Dense Layer 2             & 64                    & ReLU                         & 2112                   \\ 
Batch Norm 2              & 64                    & -                            & 256                    \\ 
Dropout 2                 & 64                    & -                            & 0                      \\ 
Dense Layer 3             & 128                   & ReLU                         & 8320                   \\ 
Dense Layer 4             & 256                   & ReLU                         & 33024                  \\ 
Output Layer              & \# transpiled params                     & -                            & 771                    \\ \hline \hline
\end{tabular}

\label{tab:decoder_model}
\end{table}

\section{Results}
\subsection{Simulation Setup}

\begin{table*}[h]
\caption{Comparison of error between the original and reverse-engineered classifiers ($i$-qubit, $j$-layer) }
\centering
\begin{tabular}{cc||ccc||ccc|c}
\multicolumn{1}{c|}{\multirow{3}{*}{\textbf{Classifier}}} & \multirow{3}{*}{\textbf{\#Params}} & \multicolumn{3}{c||}{\textbf{Brute Force \cite{ghosh2024quantumimitationgamereverse}}} & \multicolumn{4}{c}{\textbf{Proposed Idea}}                                                              \\ \cline{3-9} 
\multicolumn{1}{c|}{}  & & \multicolumn{2}{c|}{\textbf{Parameters}} & \multirow{2}{*}{\textbf{Acc.Error \%}} & \multicolumn{2}{c|}{\textbf{Parameters}}   & \multirow{2}{*}{\textbf{Acc.Error \%}} & \multirow{2}{*}{\textbf{Diff. Acc. \%}}\\
\cline{3-4} \cline{6-7}
\multicolumn{1}{c|}{}  & & \multicolumn{1}{c|}{\textbf{Mean}} & \multicolumn{1}{c|}{\textbf{Std. Dev.}} &  & \multicolumn{1}{c|}{\textbf{Mean}} & \multicolumn{1}{c|}{\textbf{Std. Dev.}} & \\
\hline \hline
1Q; 3-layer & 4 & 5.94e-02 & 8.55e-02 & 1e-16 & 3.06e-01 & 1.55e-01 & 10.8 & 8.66e-02\\ 
\hline
2Q; 1-layer & 6 & 5.33e-02 & 2.50e-02 & 1.7 & 1.19e-01 & 4.55e-02 & 7.1 & 9.03e-02\\
2Q; 2-layer & 12 & 6.10e-02 & 4.43e-02 & 3.2 & 3.17e-01 & 1.33e-01 & 11.3 & 2.9($\uparrow$)\\
2Q; 3-layer & 18 & 8.45e-02 & 8.99e-02 &  5.7 & 2.79e-01 & 1.47e-01 & 4.7 & 4.3($\uparrow$)\\
\hline
4Q; 1-layer & 8 & 7.29e-02 & 7.73e-02 & 2.1 & 5.88e-01 & 2.76e-01 & 8.2 & 4.6\\
4Q; 2-layer & 16 & 9.29e-02 & 9.91e-02 & 5.9 & 6.41e-01 & 2.03e-01 & 4.2 & 7.2($\uparrow$)\\
4Q; 3-layer & 24 & 1.18e-01 & 9.79e-02 & 6.3 & 3.77e-01 & 1.15e-01 & 3.7 & 12.1($\uparrow$)\\
\hline
8Q; 1-layer & 16 & 6.16e-02 & 3.84e-02 & 4.1 & 6.01e-01 & 2.11e-01 & 4.3 & 2.35e-02\\
8Q; 2-layer & 32 & 8.71e-02 & 3.69e-02 & 5.3 & 6.33e-01 & 1.56e-01 & 8.5 & 1.46\\
8Q; 3-layer & 48 & 1.71e-01 & 2.81e-01 & 7.6 & 4.77e-01 & 1.05e-01 & 11.9 & 4.07\\
\hline \hline
\end{tabular}
\label{table:error}
\end{table*}

We tested the proposed autoencoder-based reverse engineering on multiple QML models to extract their parameters. The QML models are implemented in Pennylane \cite{bergholm2022pennylaneautomaticdifferentiationhybrid} to utilize the \texttt{lightning.qubit} feature for performing linear algebra calculations faster. All QML models have been trained using the Gradient Descent Optimizer with a learning rate of 0.05, and a Mean Squared Error loss function has been used to evaluate the performance on the MNIST \cite{mnist} dataset picking labels as per the capacity of the QML model. The transpilation of the circuits for the QML models has been done using the transpiler library of Qiskit \cite{qiskit2024} keeping a linear coupling map, and a basis gate set of \texttt{[id, x, sx, cnot, rz]}. The autoencoder has been implemented in TensorFlow \cite{tensorflow2015-whitepaper} and the RE of the transpiled circuits to extract the parameters has been done on the same setup as the transpilation procedure, running on a machine with 16GB RAM on an Intel Core i7-6700 CPU at a clock frequency of 3.40 GHz. 

\subsection{Performance Analysis}
\begin{figure}
    
    \centering
    \includegraphics[width=1\linewidth]{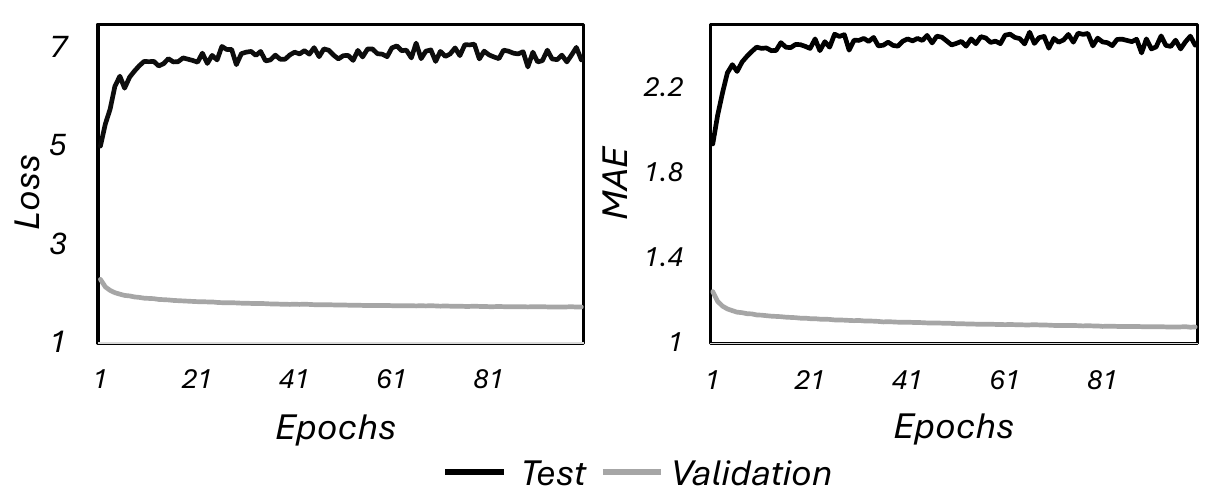}
    \caption{The performance of the autoencoder in predicting the parameters to reverse engineer the QML model. Due to the high non-linearity of the transpilation procedure, the rate of decrease of the loss is low.}
    \label{fig:autoplot}
    \vspace{-10pt}
\end{figure}

The performance of the autoencoder model is shown in Fig. \ref{fig:autoplot}. Since the model performs a regression task, we use Mean Absolute Error (MAE) to measure the accuracy of the model. We can define a positive performance of the model based on the decreasing trend of the loss as well as the MAE value. To perform an apple-to-apple comparison with the existing brute-force approach \cite{ghosh2024quantumimitationgamereverse}, we implemented their study and executed our idea against the same set of circuits. We keep the same metric for the evaluation of the efficacy of the proposed idea by comparing the percentage of decrease in the accuracy after reverse engineering the model. We compile the results in Table \ref{table:error}, and observe a minor increase in the mean error of the predicted parameters as well as the accuracy error. This is explained by the fact that transpilation is a highly non-linear mathematical procedure and hence cannot be effectively learned by the autoencoder to the level of producing extremely effective predictions of the reverse-engineered parameters from the transpiled circuit. In spite of this, the autoencoder produces parameters that are close enough for the adversary to mimic the user-designed model. 

The adversary can obtain the RE'ed model and perform a few epochs of training to reduce the difference in accuracy with the original model which presents itself as a greater threat. We train the reverse-engineered circuit for 30 epochs and present the results in the $Diff. Acc.\%$ column of Table \ref{table:error} to validate our claim. We can observe that for all classifiers, the difference in accuracy between the original model and the RE'ed model has reduced. The accuracy of the RE'ed model has also increased by a significant margin for a few classifiers (values marked with $\uparrow$). Therefore, it is safe to claim that an adversary reverse engineering any QML model is a major threat that warrants strong defense mechanisms.

\begin{figure*}
    
    \centering
    \includegraphics[width=0.9\linewidth]{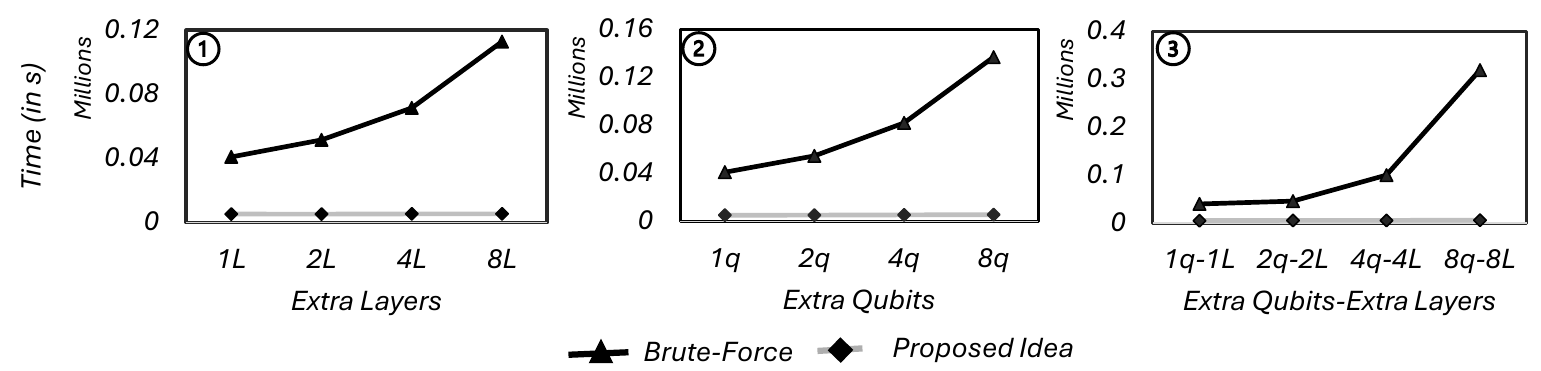}
    \caption{Performance of the proposed idea against existing countermeasures presented in \cite{ghosh2024quantumimitationgamereverse}. A 2-qubit 3-layered classifier has been reverse-engineered for this experiment. It is observed from the plots that the RE overhead is significantly reduced undermining the security of the countermeasures completely.}
    \label{fig:ctrplt}
    \vspace{-10pt}
\end{figure*}

\subsection{Overhead Analysis}
The proposed idea exhibits a significant improvement in RE overhead of QML circuits compared to \cite{ghosh2024quantumimitationgamereverse}. The dataset preparation for the autoencoder takes minimal time (of the order $10^3$ seconds) and training time is of the order $10^2$ seconds. In the reverse engineering process, parsing the circuit is done in linear time in the length of the circuit, and inferencing from the autoencoder is performed in constant time, thus reducing the overall time taken by the adversary. For example, the authors in \cite{ghosh2024quantumimitationgamereverse} find the time taken to reverse engineer a 4-qubit classifier with 16 layers to be greater than $10^7$, but the proposed approach takes time in the order of $10^2$ seconds ($10^3$ with the time for dataset preparation and training time of the autoencoder). We also observe the difference in time taken to reverse engineer a 4-qubit classifier from Table \ref{tab:layer}. For deeper quantum circuits the proposed idea provides a $\sim10^3-10^4 \times$ improvement over the brute-force approach. 
\begin{table}
\caption{Comparison of the increase in time taken to RE a 4-qubit classifier on increasing the layers}
\centering
\begin{tabular}{c||c || c}
\textbf{\# Layers} & \textbf{Brute Force \cite{ghosh2024quantumimitationgamereverse}} & \textbf{Proposed Idea}\\ \hline \hline
1  & 1.51e+04s  & 3.96e+03s \\ 
2 & 2.81e+04s  & 3.97e+03s \\ 
4 & 7.75e+05s & 3.97e+03s \\  
8 & $>$ 1e+06s & 3.99e+03s \\
16 & $>$ 1e+07s & 4.01e+03s \\ \hline \hline
\end{tabular}
\label{tab:layer}
    \vspace{-10pt}
\end{table}

\subsection{Resilience to Countermeasures}
We evaluated the effectiveness of the proposed approach against several countermeasures against the RE attack \cite{ghosh2024quantumimitationgamereverse} such as the inclusion of dummy qubits and layers to the original QML model. 
Although a brute-force approach to extracting the parameters becomes extremely slow upon implementing such countermeasures, our proposed idea is to reverse engineer circuits with extra qubits and layers in much less time. From the results in Fig. \ref{fig:ctrplt}, we can validate the claim of our proposed idea being considerably faster. For the best-suggested countermeasure (adding both dummy qubits and extra layers), our proposed method performs $10^3 \times$ better than the brute force approach for a 2-qubit 3-layered classifier with 4 dummy qubits and 4 extra layers added as a countermeasure.

\subsection{Considerations for Noise}
All QMLs were trained and inferenced under noiseless conditions. Although inherent noise in quantum hardware can affect the parameterized rotation gates during the training of a QML model, the attack model presumes that the adversary already has access to the fully trained model and then performs reverse engineering (RE) to extract the trained parameters. Consequently, the RE process is not influenced by hardware noise, making the use of noiseless simulations irrelevant to the validity of the RE concept or the overhead analysis.

\section{Conclusion}
In this paper, we discuss the vulnerability associated with Quantum Machine Learning (QML) models to reverse engineering (RE) attacks in untrusted quantum cloud environments. We proposed a novel autoencoder-based approach for extracting the trained parameters of QML models from their transpiled quantum circuits. We demonstrated that the reverse-engineered models retained significant accuracy, posing a serious threat to the integrity and confidentiality of QML systems. Through extensive testing on multi-qubit classifiers, we showed that the proposed method significantly reduced the time and computational overhead of reverse engineering compared to the state-of-the-art. 
Our findings underscore the need for developing secure quantum computing practices, particularly with third-party quantum cloud providers, to protect intellectual property and prevent unauthorized access as quantum computing becomes more widely adopted.

\section*{Acknowledgment}

The work is supported in parts by the National Science Foundation (NSF) (CNS-1722557, CCF-1718474, OIA-2040667, DGE-1723687, and DGE-1821766) and gifts from Intel.

\bibliographystyle{unsrt}
\bibliography{refs}

\end{document}